\documentclass[manuscript,authorversion,review=false,timestamp=false]{acmart}

\usepackage{balance}
\usepackage{makecell}
\usepackage{graphicx}

\AtBeginDocument{%
  \providecommand\BibTeX{{%
    \normalfont B\kern-0.5em{\scshape i\kern-0.25em b}\kern-0.8em\TeX}}}

\makeatletter
\let\@authorsaddresses\@empty
\makeatother

\begin{document}

\title{User Characteristics in Explainable AI: The Rabbit Hole of Personalization?}

\author{Robert Nimmo}
\orcid{0009-0006-5857-6469}
\affiliation{%
  \institution{University of Glasgow}
  \city{Glasgow}
  \country{United Kingdom}
}
\email{robertnimmo26@live.co.uk}

\author{Marios Constantinides}
\orcid{0000-0003-1454-0641}
\affiliation{%
  \institution{Nokia Bell Labs}
  \city{Cambridge}
  \country{United Kingdom}
}
\email{marios.constantinides@nokia-bell-labs.com}

\author{Ke Zhou}
\orcid{0000-0001-7177-9152}
\affiliation{%
  \institution{Nokia Bell Labs}
  \city{Cambridge}
  \country{United Kingdom}
}
\email{ke.zhou@nokia-bell-labs.com}

\author{Daniele Quercia}
\orcid{0000-0001-9461-5804}
\affiliation{%
  \institution{Nokia Bell Labs}
  \city{Cambridge}
  \country{United Kingdom}
}
\email{daniele.quercia@nokia-bell-labs.com}

\author{Simone Stumpf}
\orcid{0000-0001-6482-1973}
\affiliation{%
  \institution{University of Glasgow}
  \city{Glasgow}
  \country{United Kingdom}
}
\email{simone.stumpf@glasgow.ac.uk}

\begin{abstract}
As Artificial Intelligence (AI) becomes ubiquitous, the need for Explainable AI (XAI) has become critical for transparency and trust among users. A significant challenge in XAI is catering to diverse users, such as data scientists, domain experts, and end-users. Recent research has started to investigate how users' characteristics impact interactions with and user experience of explanations, with a view to personalizing XAI. However, are we heading down a rabbit hole by focusing on unimportant details? Our research aimed to investigate how user characteristics are related to using, understanding, and trusting an AI system that provides explanations. Our empirical study with 149 participants who interacted with an XAI system that flagged inappropriate comments showed that very few user characteristics mattered; only age and the personality trait openness influenced actual understanding. Our work provides evidence to reorient user-focused XAI research and question the pursuit of personalized XAI based on fine-grained user characteristics.
\smallskip

\noindent\textbf{\underline{Disclaimer}: This paper contains examples of language that some people may find offensive.}
\end{abstract}



\maketitle

\balance

\section{Introduction}
\label{sec:introductions}
Artificial Intelligence (AI) applications are increasingly becoming ubiquitous \cite{abdul_trends_2018}, from identifying spam emails to determining eligibility for loans to generating language. With the rapid advancements of AI and the seamless integration into people's lives, the need for Explainable AI (XAI) has become critical to foster transparency and trust~\cite{gunning_xaiexplainable_2019}. This need for transparency and trust is more than evident not least because of the potential repercussions of automated decisions on people's lives and the increasing regulatory oversight of AI systems~\cite{kissinger_age_2021,barredo_arrieta_explainable_2020,tahaei_human-centered_2023}.

To respond to this need, the field of XAI has made substantial progress in developing methods for providing explanations, such as global and local explanations for AI models \cite{ribeiro_why_2016, lundberg_unified_2017}. Global explanations offer an overview of the model's functioning, whereas local explanations delve into how specific predictions were generated for individual instances. Explanations in AI systems can assist end users in constructing accurate mental models of AI systems, thereby enhancing both their understanding and the degree of trust attributed to these systems~\cite{lim_why_2009,kulesza_too_2013}. These methods have enhanced users' comprehension of AI systems by `opening up' their black-box inner workings \cite{gunning_xaiexplainable_2019}.

At the same time, there is growing evidence suggesting that a user's characteristics may have an impact on how explanations are received~\cite{ehsan_who_2021, jiang_who_2022, stumpf_integrating_2008}. A user's characteristics typically consist of a broad range of attributes, including but not limited to age, gender, personality, and experience. The influence of user characteristics has been studied in other fields, for example, personality has been extensively studied in the context of recommender systems~\cite{tintarev_survey_2007,tkalcic_personality_2015, dhelim_survey_2022} and Human-Robot Interaction~\cite{craenen_shaping_2018,craenen_we_2018}. Information retrieval research has revealed that previous experience impacts search behavior \cite{yan_concept-based_2006, ranganathan_mashup-based_2009, white_characterizing_2009, soulier_domain_2014, mao_how_2018}. Specific to XAI, gender and educational background have been found to affect the types of explanations that participants preferred as well as their trust \cite{reeder_evaluating_2023}. Personality traits have been found to impact the interaction with explanations \cite{millecamp_explain_2019,millecamp_whats_2020}. This has provided some impetus to work that aims to employ users' characteristics to personalize  XAI and tailor the design and presentation of explanations to users \cite{conati_toward_2021}.

While these works provide some evidence about a potential link between user characteristics and AI explanations, our understanding remains limited. Therefore, we set out to test whether user characteristics are associated with user engagement with explanations, the understanding of these explanations, and the degree of trust users attribute to AI systems. In so doing, we developed an interactive AI system that provides local explanations for flagging inappropriate comments, and formulated four Research Questions (RQs):

\begin{enumerate}
    \item[\textbf{RQ1}:] How do user characteristics relate to user engagement in finding misclassifications? 
    \item[\textbf{RQ2:}] How do user characteristics relate to the perceived understanding of explanations?
    \item[\textbf{RQ3:}] How do user characteristics relate to the actual understanding of explanations?
    \item[\textbf{RQ4:}] How do user characteristics relate to the degree of trust users attribute to an AI system?
\end{enumerate}

In addressing these questions, we make three main contributions: 

\begin{itemize}
\item We provide a prototype that provides local explanations in the task of flagging inappropriate comments (section \ref{sec:prototype}), which we release open-source. The prototype guides participants through the user study and logs users' interactions. We utilized a pre-existing multi-class classifier named Detoxify \cite{hanu_detoxify_2020} with our prototype, where we generated explanations for each comment. Participants interacted and provided feedback on the explanations provided using the prototype.

\item We advance the methodology of investigating user characteristics in XAI. We conducted a large-scale study by engaging 149 participants recruited from Prolific (section \ref{sec:study}). We utilized our prototype to investigate the effect users' characteristics - specifically age, gender, previous experience, and the Big 5 personality traits - have on outcome measures, such as participants' engagement, trust, and perceived and actual understanding of the XAI system.

\item We show empirical evidence that very few user characteristics are related to our outcome measures of engagement, trust, and understanding. Only age and the personality trait openness were associated with actual understanding (section \ref{sec:results}).

\end{itemize}

In light of these results, we argue to reorient user-focused XAI research in pursuit of personalized XAI based on fine-grained user characteristics (section \ref{sec:discussion}).

\section{Related Work}
\label{sec:related}

\subsection{Transparency, Understanding, and Explanations}

Many AI systems are difficult to understand as they typically are `black-box models'~\cite{adadi_peeking_2018}. The need for transparent AI systems has become increasingly recognized \cite{barredo_arrieta_explainable_2020} in the context of responsible AI systems where explanations are commonly employed to increase the transparency of AI systems.  It has been suggested that explanations help to evaluate if an AI system meets fairness, privacy, reliability, causality, and trust which can help to expose potential issues in an AI system during development before being deployed to the end-user \cite{doshi-velez_towards_2017}. Explanations can also help end-users understand how the underlying AI model works \cite{lim_why_2009,kulesza_too_2013}, leading to higher user satisfaction and better-calibrated reliance on AI systems \cite{dzindolet_role_2003,cramer_effects_2008}, reducing the potential of algorithm aversion \cite{de-arteaga_case_2020}, obtaining better feedback and input to an AI system \cite{kulesza_principles_2015} and increasing trust \cite{el_bekri_study_2020, barredo_arrieta_explainable_2020, petch_opening_2022}. 

Explanations are designed to help the user understand how the AI model behaves, by providing context on how and why a specific decision was made (local explanation) or by explaining how the model works (global explanation)  \cite{gunning_xaiexplainable_2019,liao_introduction_2021, dodge_explaining_2019, ehsan_operationalizing_2021}. Explanations frequently target users' \emph{mental models} \cite{norman_observations_1987, kulesza_too_2013} which are an individual's internal representation or model of how a system works, which gets updated every time the person interacts with an AI system. Mental models assist individuals in understanding, explaining, and predicting events and determining appropriate actions. Lack of transparency does not prevent users from building a mental model, although it can lead to an inaccurate mental model being built. When providing explanations, it has been suggested that they need to cover a set of `intelligibility types' \cite{lim_assessing_2009, kulesza_why-oriented_2011} which flesh out the user's mental model, addressing \emph{Why} (why did the AI system do X?), \emph{Why not} (why did the AI system not do Y?), \emph{What if} (what would the AI system do if Z happens?), \emph{How} (how does the AI system do Y?), and \emph{What} (what did the AI system do?).

Two of the most popular explanation approaches, addressing the \textit{Why} aspect of explanations, include SHapley Additive exPlanations (SHAP) \cite{lundberg_unified_2017} and Local Interpretable Model-Agnostic Explanations (LIME) \cite{ribeiro_why_2016}. Both approaches are model-agnostic approaches, allowing them to work for any model. SHAP explanations provide a local and global explanation of the given model. SHAP is an extension of Shapley Values \cite{shapley_notes_1951}, a solution from cooperative game theory. SHAP explanations are more complex and thus more difficult to understand than LIME explanations; many applications using SHAP are targeted at AI experts. Instead, LIME provides a local explanation of the given model by showing the most important features used in making the decision; they have shown good usability and interpretability for a range of users. 

When developing explanations, there is no ``one-fits-all approach'' \cite{liao_human-centered_2022, gunning_xaiexplainable_2019}. Previous research has indicated that current XAI systems are often designed based on the researcher's or designer's subjective perception of what constitutes a `good' explanation rather than being tailored to the needs of the end-user \cite{miller_explainable_2017, miller_explanation_2019}. Instead, explanations should be developed based on the needs of the users and the purpose of the explanation through a human-centered approach, being specific to `who' the explanation will be seen by and `why' they require the explanation~\cite{ehsan_human-centered_2020}. Consequently, a number of research efforts have tried to uncover user characteristics that matter in interacting with explanations, with a view to personalizing explanations to a specific user.

\subsection{User Characteristics and Personalization}
User characteristics have frequently been used to personalize experiences, interactions, or recommendations. A new direction is to use the same principle of personalization for XAI, in which the design and presentation of explanations are tailored to users \cite{conati_toward_2021}. 

A user can have many different user characteristics, such as age, gender, previous experience, and personality. We present an overview of how these user characteristics have been employed in research towards personalization. It has been found that \emph{gender} can influence the perception and experience of using recommender systems \cite{knijnenburg_explaining_2012}. Gender and educational background have also been found to affect the types of explanations that participants preferred as well as their trust \cite{reeder_evaluating_2023}. However, it has also been cautioned that gender might not be as important as previously thought and requires more research to confirm any effect \cite{tran_user_2023}. A common user characteristic that is captured during research studies is \emph{age}. Although not proven to be significant, age has been considered in the willingness to accept advice \cite{jiang_who_2022}. \emph{Previous experience} has been found to influence search behavior for information \cite{yan_concept-based_2006, ranganathan_mashup-based_2009,white_characterizing_2009,soulier_domain_2014,mao_how_2018}, how interactions occur with recommendations in a recommender system \cite{cai_impacts_2022} and also preferences for the presentation format of explanations and understanding of explanations \cite{szymanski_visual_2021}. However, other research has revealed no bearing of previous experience with AI and with interaction with an XAI system \cite{leichtmann_effects_2023}.

\emph{Personality} has recently emerged as an area of focus for research in personalizing AI or XAI. In recommender systems, personality has been used to suggest appropriate items where there is little information about a user or an item to give meaningful recommendations \cite{tintarev_adapting_2013,tkalcic_personality_2015, liao_user_2022}. In addition, personality has also been used in Human-Robot Interaction (HRI) research to improve users' perception and interactions with robots \cite{craenen_we_2018, craenen_shaping_2018}. This includes looking at the role personality plays in AI anxiety \cite{kaya_roles_2022} and how personality can affect perception and trust in recommender systems recommendations \cite{cai_impacts_2022}. Little research has been conducted to investigate how personality affects XAI but personality traits have been found to impact the interaction with explanations \cite{millecamp_explain_2019,millecamp_whats_2020, conati_toward_2021}. 

\subsection{Research Gap}

Previous research has indicated that user characteristics might impact various aspects of system use. Considering the growing importance of AI systems and the imperative to provide explanations, there have been attempts to investigate how user characteristics influence users' perception and understanding of such explanations. However, our understanding of whether and how different user characteristics are associated with the effectiveness of AI explanations remains limited. Therefore, investigating this can potentially improve explanation design and lead to the creation of personalized explanations.

\section{Methods}
\label{sec:methods}
To investigate our research questions, we ran an empirical, online study in which 149 participants interacted with an AI prototype that classified Internet comments as `toxic' or `non-toxic' for flagging inappropriate comments for humans to check. This is a commonplace task, similar to other binary classification systems such as spam filters, where human-in-the-loop input is beneficial. The interface offered explanations and the ability to make suggestions as to how to improve the system, inspired by the explanatory debugging approach~\cite{kulesza_principles_2015}. We collected the participants' characteristics as well as prototype usage information, understanding and trust for statistical analyses.

\subsection{Prototype}
\label{sec:prototype}
\subsubsection{The Dataset}

We decided to use a hate speech dataset, Jigsaw  \cite{cjadams_toxic_2017}, for our study. This dataset is freely available and several open-source pre-trained models have been developed on this dataset. In addition, the dataset does not require prior knowledge or domain specialization in a particular field so is ideally suited for user studies with lay people. 

It contains 220,000 Wikipedia talk page edit comments over one year, which have been labeled by human raters for toxic behavior in six categories. These categories included \emph{toxic}, \emph{severe toxic}, \emph{obscene}, \emph{threat}, \emph{insult}, and \emph{identity hate}. For our study, we filtered out comments with high (>0.75) obscene or identity hate scores in order to mitigate harm to our participants. With the filtered toxic/non-toxic Jigsaw test dataset, we randomly sampled 200 comments while keeping an equal class distribution between toxic and non-toxic comments, using the Python library Pandas built-in function. Further, we removed any comments that were too short (if the comment was less than three words), not understandable (if the comment is written in a language other than English or containing numerous misspellings, rendering the comment illegible), or inappropriate (if the comment fell into removed categories through a manual review to satisfy ethics requirements). This resulted in 100 comments which were shown to the study participants. 

\subsubsection{The AI model}

We employed a pre-existing AI classifier named Detoxify \cite{hanu_detoxify_2020}. This model is a multi-class classifier and was trained on the Jigsaw training dataset using a BERT base model (uncased) transformer from Hugging Face \cite{devlin_bert_2019}. During each training step, a batch size of 10 was utilized. For each input, the data was tokenized using the Bert tokenizer. A binary cross-entropy loss function was used to optimize the model between each step. We selected this model as it was open-source and pre-trained, allowing us to quickly implement the model in our prototype and develop explanations. The model achieved an AUC score of 98 \cite{hanu_detoxify_2020}; thus, it still produces misclassification to engage our participants. Our research focus concentrated on reactions to explanations instead of model accuracy; this also influenced our choice of using a state-of-the-art classifier to minimize the impact of lower model accuracy on measures under investigation, such as trust.

\subsubsection{Explanations}
We decided to use Local Interpretable Model-Agnostic Explanations (LIME) \cite{ribeiro_why_2016} to explain our model. This is an approach that has been frequently employed to explain deep learning models such as the one we employed in our study. 

To generate our explanations, we used the Python implementation of LIME \cite{ribeiro_lime_2023}. The LimeTextExplainer was used to explain our predictions for each comment in our testing dataset. The LIME explanation generated a list of the top 10 most important words and their weights for each comment, which were stored in the prototype. Table \ref{tab:top-10-example} provides an example of the LIME output for the top 10 most significant words generated by Lime. The higher the word weight, the more significant the word was to the prediction. These explanations were saved in a JSON file to be used in our prototype.

\setlength{\tabcolsep}{2.75em}
\begin{table}
\centering
\caption{Top 10 most important words for the comment ``Ah, the wonders of political correctness. Bastard, a perfectly good word for centuries, is no longer to be used. Where do you suppose it will end?''}
\label{tab:top-10-example}
\begin{tabular}{@{}lll@{}}
\toprule
\multicolumn{1}{c}{Word} & \multicolumn{1}{c}{Weight} & Label \\ \midrule
Bastard & .829 & Toxic \\
perfectly & .030 & Toxic \\
the & .021 & Non-toxic \\
Where & .020 & Toxic \\
good & .016 & Non-toxic \\
Ah & .013 & Non-toxic \\
political & .012 & Non-toxic \\
is & .011 & Non-toxic \\
centuries & .008 & Non-toxic \\
used & .002 & Toxic \\ \bottomrule
\end{tabular}

\end{table}
\setlength{\tabcolsep}{1em}

\subsubsection{Interface}

\begin{figure*}
\begin{center}
\includegraphics[scale=0.21]{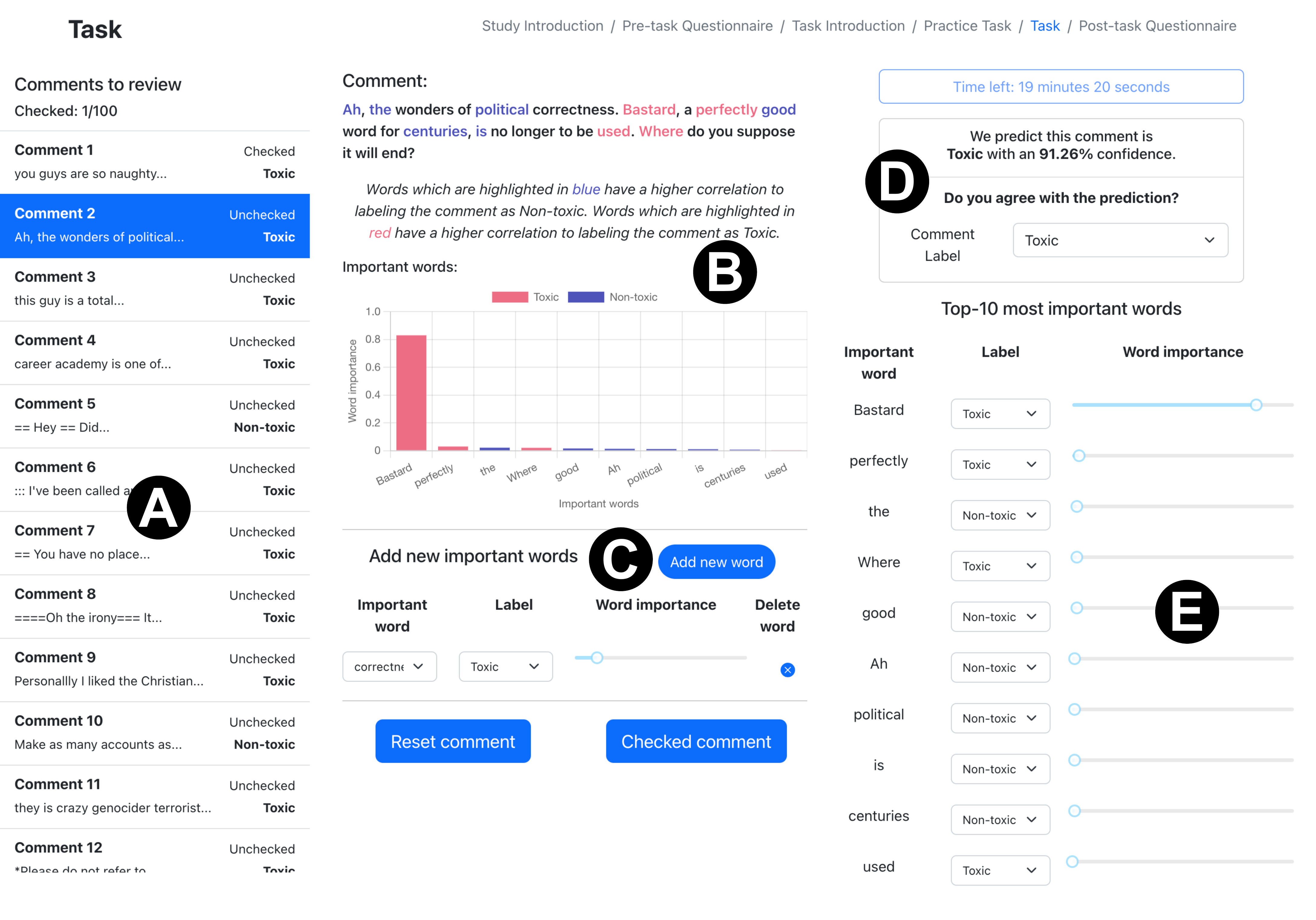}
\end{center}
\caption{\label{fig:prototype_diagram}The prototype interface. (A) A list of all the comments to be assessed. (B) The selected comment and a bar chart representing the top 10 most important words in the selected comment are sorted in descending order. (C) Ability to add new important words that are not already highlighted. (D) A summary of the AI system's prediction for the selected comment and the ability to change the predicted label. (E) A list of the top 10 most important words in the selected comment. For each important word, the user can change the label (which label, Toxic/Non-toxic, is most associated with the word) and the word importance (how important the word is in the prediction).}
\Description{The prototype interface: The interface includes on the far left a list of all the comments to be assessed. On clicking a comment in this list, the selected comment text is shown to the right of the list, with words having a higher correlation with the non-toxic label highlighted in blue, or in red for the toxic label. A bar chart representing the top 10 most important words in the selected comment are shown below the comment sorted in descending order, again color-coded in blue or red to show the relevance to the label. Underneath the bar chart, the user has the ability to add new important words that are not already highlighted by selecting them from a drop-down list, adding the label, and using a slider to give the importance of this word. To the far right of the interface is a summary of the AI system's prediction for the selected comment and the ability to change the predicted label by selecting it through a drop-down. Underneath this is a list of the top 10 most important words in the selected comment (similar to the bar chart) but for each important word, the user can change the label through a drop-down and the word importance through a slider.}
\end{figure*}

Inspired by work on explanatory debugging \cite{kulesza_principles_2015, stumpf_interacting_2009, kulesza_fixing_2009, kulesza_explanatory_2010}, we decided to develop an interface that might be used in a task to review automatically flagged comments (Figure \ref{fig:prototype_diagram}). 

\sloppy
The interface is comprised of a list of comments to check (\textbf{\textit{Component A}}), and its predicted class. Based on the user's selection, the comment text is displayed next to it (\textbf{\textit{Component B}}), with the top 10 most important words highlighted in the comment, color-coded according to class. Underneath the comment text, it is displayed a bar chart to visualize the weight each important word had on the prediction. To the right of the comment text, we showed the predicted class and the Detoxify model prediction confidence. The user is able to agree with this class prediction or assign a different label (\textbf{\textit{Component D}}). To provide additional feedback to the system, the user is able to add new words not highlighted in the comment (\textbf{\textit{Component C}}), or adjust the weight and class of words in the comment (\textbf{\textit{Component E}}). The user can reset any changes made, or mark the comment as checked. Comments were presented in the same order to all participants. Given the purpose of the study was to investigate whether any user characteristics are associated with users' perception and understanding of explanations, this choice allowed us to eliminate any confounding factors that might have been introduced by varying the order of the comments.

\subsubsection{Prototype Implementation}
We developed the prototype as a web app, accessible through a URL. The prototype is modular and reusable in the future for conducting similar user studies. It utilizes serverless functions, specifically Cloudflare Worker Functions, a MongoDB database as the backend and database, and a React web app frontend. Using serverless functions allows future studies to decide the specific backend architecture language they would like to use, as all requests to the backend are completed through API requests. In addition, serverless functions are quicker, easier, and cheaper to deploy and require less setup during the deployment configuration. Lastly, they are scalable, being able to cope with a large number of participants in a short time span. MongoDB was selected due to its flexibility, scalability, and ease of use, being more suitable for storing long user logs with its document-oriented approach that closely matched the collected user log data structure. The React web app frontend was chosen for its flexibility and ease of development. React's component-based architecture allowed us to easily modularise the web apps, allowing the prototype to be customized easily for future user studies. It also allowed us to build a responsive and interactive user interface, which is crucial for a good user experience.

\subsection{Participants}

To determine the required sample size for our analyses, we used G*Power.\footnote{\url{https://www.psychologie.hhu.de/arbeitsgruppen/allgemeine-psychologie-und-arbeitspsychologie/gpower}} We found that at least 117 participants were required to analyze our independent variables (i.e., age, gender, previous experience with AI, and the five personality traits) using two-tailed regressions at a minimum of 0.05 level of significance, 95\% level of power, and an effect size of 0.2. To account for any data loss due to quality checks or missing data, we further increased our sample size. By conducting a post-hoc sensitivity analysis, we found that our analyses had even more strict effect size of 0.15.

We thus recruited 150 participants for our user study through Prolific (https://www.prolific.co/), an online platform to recruit participants for research studies. One participant had incomplete demographic information and was therefore excluded from the analysis, thus resulting in 149 participants in our data set. 

Our participants' ages ranged from 18 to 67, with a median age of 36 ($\sigma=11.5$). There were 75 male participants and 74 female participants in total. Participants needed to be fluent in English in order to judge the toxicity of language in a comment and we enforced this through a screener question and by restricting the residence requirement to majority-English speaking countries. In keeping with our use case of moderating comments, we did not restrict participants to have AI knowledge.

Participants were compensated £4.50, consistent with Prolific's payment principles of `ethical rewards' and in line with the United Kingdom national minimum wage at the time the study was conducted. We obtained informed consent; our user study was considered and approved by the University of Glasgow School of Computing Science Research Ethics Committee. 

\subsection{Study Task}
\label{sec:study}
\begin{figure*}
\begin{center}
\includegraphics[width=\linewidth]{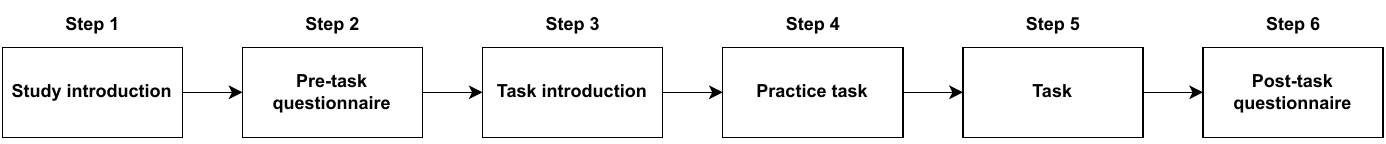}
\end{center}
\caption{\label{fig:user_study_procedure}
Our user study procedure consists of six steps. \textit{Step 1} - Study introduction: The participants were provided with an overview of the study's objectives, the potential outcomes of their involvement, the utilization of their data, as well as the advantages and disadvantages of participation. \textit{Step 2} - Pre-task questionnaire: The participants completed a pre-task questionnaire consisting of two sections to measure their personality and previous experience with AI systems. \textit{Step 3} - Task introduction: Participants were presented with a scenario and an introduction to the task that they were required to complete for the prototype. \textit{Step 4} - Practice task: Participants were instructed to complete a practice task using the explanatory debugging interface. \textit{Step 5} - Main task: The primary task involved the utilization of the explanatory debugging interface to complete the study task. \textit{Step 6} - Post-task questionnaire: Participants were requested to fill out a post-task questionnaire to measure the participant's trust, perceived understanding, and actual understanding.}
\Description{Our user study procedure consists of six consecutive steps: Step 1. Study introduction: The participants were provided with an overview of the study's objectives, the potential outcomes of their involvement, the utilization of their data, as well as the advantages and disadvantages of participation. Step 2. Pre-task questionnaire: The participants completed a pre-task questionnaire consisting of two sections to measure their personality and previous experience with AI systems. Step 3. Task introduction: Participants were presented with a scenario and an introduction to the task that they were required to complete for the prototype. Step 4. Practice task: Participants were instructed to complete a practice task using the explanatory debugging interface. Step 5. Main task: The primary task involved the utilization of the explanatory debugging interface to complete the study task. Step 6.  Post-task questionnaire: Participants were requested to fill out a post-task questionnaire to measure the participant's trust, perceived understanding, and actual understanding.}
\end{figure*}

After participants agreed to the HIT on Prolific, they were directed to our study site. The user study consisted of six steps summarized by Figure \ref{fig:user_study_procedure}, which took about 30 minutes to complete.

We first familiarized participants with the study and obtained informed consent. We then presented them with a pre-task questionnaire consisting of two sections. In the first section, we captured a participant's personality traits using TIPI \cite{gosling_very_2003}. The second section employed the level of expertise scale \cite{mcbeath_levels_2011} for, respectively, AI systems, machine learning, and explanations. After this, participants were introduced to their task: \emph{``You are currently working for a social media platform like Reddit. The company has just deployed a new AI system that automatically flags a comment as Toxic or Non-toxic.''}, and given instructions on how to use the prototype interface. In order to facilitate familiarity and proficiency in using the interface, participants were instructed to complete a practice task. The interface for the practice task mirrored that of the main task, but it only featured one comment to check and incorporated two explicit attention checks, which each participant was asked to complete. Participants were then directed to the main task interface (Figure \ref{fig:prototype_diagram}). The prototype presented participants with 100 comments to check, with 61 toxic comments and 39 non-toxic comments for our study's final class distribution. Each participant was allotted 20 minutes to complete the main study task, which entailed evaluating as many of the 100 comments supplied as possible. Participants were not expected to examine every comment but rather to spend time to thoroughly think through each comment. Once the time elapsed they were directed to the post-task questionnaire. There, participants were asked about their trust in the system using the trust scale recommended for XAI \cite{hoffman_metrics_2019} and rated their perceived level of understanding regarding the AI system. They also completed a series of 12 questions to ascertain their actual level of understanding,  similar to previous work by Kulesza et al.~\cite{kulesza_tell_2012} to measure the mental model score. These questions covered nine key criteria regarding the AI system. These criteria included understanding of the user interface, feature weights, feature labels, adjusting feature weights, classes, adding new words, confidence, how predictions are made, and what class might be predicted based on features. Nine of the questions were multiple-choice, while three questions allowed for open-ended answers.

\subsection{Data Collection and Analysis}
One of the most common ways to characterize personality is the five-factor model of personality (Big 5) \cite{mccrae_validation_1987}, where personality is measured on five dimensions: 
\begin{itemize}
    \item Openness: measures how curious, original, and open to new ideas an individual is.
    \item Conscientiousness: measures how self-disciplined and hardworking an individual is.
    \item Extraversion: measures how sociable, outgoing, and assertive an individual is.
    \item Agreeableness: measures how cooperative and good-natured an individual is.
    \item Neuroticism: measures an individual tendency towards unstable emotions.
\end{itemize}

Many questionnaires have been developed of different lengths and accuracy to measure a participant's personality against the Big 5. One of the most well-established and used methods is the Big Five Inventory (BFI) \cite{john_big_1991,benet-martinez_cinco_1998,john_paradigm_2008}. The BFI is a 44-item questionnaire which consists of participants answering each question with a 5-point Likert scale. However, in studies where time is limited, the Ten-Item Personality Inventory (TIPI) of measuring a participant's personality against the Big 5 has been used \cite{gosling_very_2003}. Although similar to BFI, it only includes ten items and utilizes a 7-point Likert scale. It has been shown that TIPI gathers similar results compared to BFI but can be less robust for the openness dimension \cite{gosling_very_2003}. In our study, we chose to employ the TIPI questionnaire. 

The responses from the pre-task questionnaire were processed as follows: we calculated participants' personality scores on the 5 dimensions using the responses from the TIPI questionnaire and the method to calculate each personality dimension set out by Gosling et al. \cite{gosling_very_2003}. The participants self-rated their level of experience with AI systems, ranging from 0 to 10. In our sample, participants previous experience ranged from 1 to 6 with a $\mu=3.15$ and $\sigma=.964$ (95\% of the values were in the range 1 to 4). While our sample skews towards lower previous experience, this reflects use cases where AI explanations are targeted at non-expert users.

From the post-task questionnaire, we obtained their system trust score, their perceived understanding score and their actual understanding score. The trust survey consisted of eight questions that assessed the participants' trust in the XAI system, utilizing a 5-point Likert scale ranging from 1 (strongly disagree) to 5 (strongly agree). To obtain the participants' overall trust score, the sum of the eight questions was calculated, with the scale reversed for question six. A participant could score a maximum of 40 (reflecting high trust in the AI system) and a minimum of 8 (reflecting low trust in the AI system). For the perceived understanding, we recorded an integer using a 5-point Likert scale that ranged from 1 (did not understand at all) to 5 (fully understood). All questionnaires used in the study were made available as supplementary material.

An automated script was used to grade the multiple-choice answers for actual understanding, awarding one mark for a correct response and zero for an incorrect answer. A researcher reviewed and graded the open-ended questionnaire responses using a predetermined marking scheme, allowing full marks for correct reasons and for half marks for partially correct reasons. The actual understanding score for each participant was calculated by summing their points, with a maximum possible score of 15.

During the main study task, we logged every user interaction with the prototype to keep track of all suggestions and changes the user made to the AI system. We ensured continued engagement and high data quality through explicit attention checks and checking the user logs for suspicious behavior, such as prolonged intervals between reviewing each comment (5+ minutes) or rapidly reviewing comments (within 1-3 seconds). 

To answer our research questions, we conducted multiple regression analyses, using user characteristics as independent variables and trust, perceived, and actual understanding as dependent variables (or outcome measures). We assessed the linearity by partial regression plots and a plot of studentized residuals against the predicted values. The independence of residuals was assessed with the Durbin-Watson statistic. We visually inspected a plot of studentized residuals against unstandardized predicted values to assess homoscedasticity in each multiple regression test. Multicollinearity was assessed for each multiple regression test by checking tolerance values were greater than 0.1. In addition, we examined studentized deleted residuals to ensure that no values deviated by more than ±3 standard deviations, checked that leverage values were below 0.2, and verified that values for Cook's distance exceeded 1. Lastly, we checked the assumption of normality for the residual was met for each of the multiple regression tests by Q-Q plots. 

\section{Results}
\label{sec:results}

We first present the statistics of our participants, and then present a correlation analysis between variables. We then show the results of our regression analysis, aiming to scrutinize the impact of user characteristics---age, gender, previous experience, and personality traits---on the four dimensions of interest. These dimensions pertain to user engagement with explanations, the perceived and actual understanding of these explanations, and the degree of trust attributed to them.

\subsection{Participants Statistics}
\label{subsec:participants_stats}
The age distribution of our participants spanned from 18 to 67 years, with a median age of 36 ($\sigma=11.5$), and the gender distribution was equally divided (74 females).

On average, participants' personality traits, on a scale from 1 to 7, were low in extraversion ($\mu=3.31$ and $\sigma=1.53$), high in agreeableness ($\mu=5.03$ and $\sigma=1.16$), high in conscientiousness ($\mu=5.07$ and $\sigma=1.35$), average in neuroticism ($\mu=4.27$ and $\sigma=1.49$) and high in openness ($\mu=4.97$ and $\sigma=1.10$); these distributions conformed to the standard personality values derived from a representative U.S. population sample~\cite{soto_age_2011}.

\subsection{Correlations Between Variables}
We conducted a correlation analysis to investigate any relationships within our collected data. Table \ref{tab:cross_correlation} shows the correlation coefficient \emph{r} in a variable matrix. We found that there were some statistically significant correlations within personality traits: extraversion was weakly positively correlated with openness (\emph{r}=0.397),  conscientiousness (\emph{r}=0.318), neuroticism (\emph{r}=0.263) and agreeableness (\emph{r}=0.173), and neuroticism was weakly positively correlated with conscientiousness (\emph{r}=0.394) and agreeableness (\emph{r}=0.180). We also found a statistically significant weak positive correlation between age and conscientiousness (\emph{r}=0.276), and a weak negative correlation between gender and agreeableness (\emph{r}=-0.264). In our regression models, we set female participants to 0 values and male participants to 1 values. Looking at relationships within other user characteristics, previous experience and gender were weakly positively correlated (\emph{r}=0.215). 

In terms of user characteristics and outcome measures, we found that actual understanding was weakly negatively correlated with openness (\emph{r}=-0.191), conscientiousness (\emph{r}=-0.198) and agreeableness (\emph{r}=-0.173), while perceived understanding was weakly positively related to extraversion \emph{r}=0.169) and trust (\emph{r}=0.204). We therefore explored the relationships between user characteristics and outcome measures more thoroughly through regression models.

\begin{table*}
\centering
\caption{Pairwise Spearman's correlations among the independent variables Big-Five personality traits, age, gender, and previous experience, and the dependent variables trust, engagement, perceived understanding, and actual understanding. The correlations that are statistically significant are in bold and are marked with a number of *s based on their significance levels (i.e., *$p<.05$, **$p<.01$, ***$p<.001$). Significant results are discussed in the text.}
\resizebox{\textwidth}{!}{%
\begin{tabular}{@{}lllllllllllll@{}}
\toprule
\textbf{Variable} & \begin{tabular}[c]{@{}l@{}}1.\\ Openness\end{tabular} & \begin{tabular}[c]{@{}l@{}}2.\\ Conscientiousness\end{tabular} & \begin{tabular}[c]{@{}l@{}}3.\\ Extraversion\end{tabular} & \begin{tabular}[c]{@{}l@{}}4.\\ Agreeableness\end{tabular} & \begin{tabular}[c]{@{}l@{}}5.\\ Neuroticism\end{tabular} & \begin{tabular}[c]{@{}l@{}}6.\\ Age\end{tabular} & \begin{tabular}[c]{@{}l@{}}7.\\ Gender\end{tabular} & \begin{tabular}[c]{@{}l@{}}8.\\ Prev Exp\end{tabular} & \begin{tabular}[c]{@{}l@{}}9.\\ Trust\end{tabular} & \begin{tabular}[c]{@{}l@{}}10.\\ Engagement\end{tabular} & \begin{tabular}[c]{@{}l@{}}11.\\ Perc Und \end{tabular} & \begin{tabular}[c]{@{}l@{}}12.\\ Act Und\end{tabular} \\ \midrule

1. Openness & - &  &  &  &  &  &  &  &  &  &  &  \\
2. Conscientiousness & 0.105 & - &  &  &  &  &  &  &  &  &  &  \\
3. Extraversion & \textbf{0.397**} & \textbf{0.318**} & - &  &  &  &  &  &  &  &  &  \\
4. Agreeableness & 0.154 & 0.114 & \textbf{0.173*} & - &  &  &  &  &  &  &  &  \\
5. Neuroticism & 0.076 & \textbf{0.394**} & \textbf{0.263**} & \textbf{0.180*} & - &  &  &  &  &  &  &  \\
6. Age & 0.021 & \textbf{0.276**} & 0.064 & 0.032 & 0.092 & - &  &  &  &  &  &  \\
7. Gender & -0.089 & -0.065 & -0.074 & \textbf{-0.264**} & 0.153 & -0.023 & - &  &  &  &  &  \\
8. Prev Exp & 0.104 & -0.075 & -0.015 & -0.141 & 0.115 & -0.074 & \textbf{0.215**} & - &  &  &  &  \\
9. Trust & -0.128 & 0.087 & 0.061 & 0.145 & -0.034 & -0.134 & 0.021 & -0.027 & - &  &  &  \\
10. Engagement & 0.054 & 0.107 & 0.061 & 0.070 & 0.018 & -0.072 & 0.019 & -0.035 & -0.088 & - &  &  \\
11. Perc Und & 0.131 & 0.085 & \textbf{0.169*} & 0.049 & 0.144 & \textbf{-0.163*} & 0.113 & 0.087 & \textbf{0.204*} & -0.100 & - &  \\
12. Act Und & \textbf{-0.191*} & \textbf{-0.198*} & -0.140 & \textbf{-0.173*} & -0.074 & \textbf{-0.240**} & -0.007 & 0.006 & 0.029 & \textbf{0.194*} & 0.014 & - \\ \bottomrule
\end{tabular}%
}

\label{tab:cross_correlation}
\end{table*}

\subsection{Impact of User Characteristics on Outcome Measures}

\begin{table*}
\centering
\caption{Multiple regression results for the impact of age, gender, and previous experience with AI on engagement, perceived understanding, actual understanding, and trust. \textit{B} = unstandardized regression coefficient; \textit{Std. error} = standard error of the coefficient; \textit{$\beta$} = standardized coefficient; \textit{$adj.R^2$} = adjusted $R^2$. Statistically significant results are in bold and are marked with a number of *s based on their significance levels (i.e., *$p<.05$, **$p<.01$, ***$p<.001$). Significant results are described in the text.} 
\begin{tabular}{@{}lllll@{}}
\toprule
\parbox[t]{2in}{\textbf{Impact of age, gender, and previous experience with AI on engagement}} & $adj.R^2=-.017$ & $p=.926$  &  &  \\
\midrule
Predictor & $B$ & Std. error & $\beta$ & $p$-value \\ 
\midrule
Constant & 5.509 & 1.142 &  &  \\
Age & -.002 & .020 & -.009 & .912 \\
Gender & .320 & .474 & .058 & .501 \\
Previous Experience & -.027 & .248 & -.009 & .913 \\
 &  &  &  &  \\
 \midrule
\parbox[t]{2.5in}{\textbf{Impact of age, gender, and previous experience with AI on perceived understanding}} & $adj.R^2=.024$ & $p=.091$  &  &  \\
\midrule
Predictor & $B$ & Std. error & $\beta$ & $p$-value \\ 
\midrule
Constant & 3.283 & .326 &  &  \\
Age & -.011 & .006 & -.152 & .065 \\
Gender & .140 & .135 & .086 & .304 \\
Previous Experience & .073 & .071 & .086 & .306 \\
 &  &  &  &  \\
 \midrule
\parbox[t]{2.5in}{\textbf{Impact of age, gender, and previous experience with AI on actual understanding}} & $adj.R^2=.056$ & $\boldsymbol{p=.010*}$  &  &  \\
\midrule
Predictor & $B$ & Std. error & $\beta$ & $p$-value \\ 
\midrule
Constant & 13.868 & 1.067 &  &  \\
Age & -.065 & .019 & -.275 & \textbf{<.001***} \\
Gender & .115 & .443 & .021 & .795 \\
Previous Experience & -.130 & .231 & -.046 & .572 \\
 &  &  &  &  \\
 \midrule
\parbox[t]{2.1in}{\textbf{Impact of age, gender, and previous experience with AI on trust}} & $adj.R^2=-.004$ & $ p=.506$  &  &  \\
\midrule
Predictor & $B$ & Std. error & $\beta$ & $p$-value \\ 
\midrule
Constant & 24.597 & 2.065 &  &  \\
Age & -.054 & .037 & -.122 & .141 \\
Gender & -.037 & .857 & -.004 & .966 \\
Previous Experience & -.219 & .448 & -.041 & .626 \\ \bottomrule
\end{tabular}
\label{tab:age-gender-prev_exp-results}
\end{table*}

Table~\ref{tab:age-gender-prev_exp-results} shows the results of multiple regressions to test the impact of age, gender, and previous experience with AI on engagement, perceived understanding, actual understanding, and trust. When we tested the association between age, gender, previous experience, and our participants' engagement with explanations, we found no statistically significant effect. For perceived understanding, our model had no overall predictive power. 

For actual understanding, however, our model did show statistical significance. Age was found to be statistically significant and negatively associated with actual understanding. 

For trust, we found no statistically significant effect, with none of age, gender, and previous experience being statistically significant.

\begin{table*}
\centering
\caption{Multiple regression results for the impact of personality traits and previous significant results on engagement, perceived understanding,  actual understanding, and trust. \textit{B} = unstandardized regression coefficient; \textit{Std. error} = standard error of the coefficient; \textit{$\beta$} = standardized coefficient; \textit{$adj.R^2$} = adjusted $R^2$. Statistically significant results are in bold and are marked with a number of *s based on their significance levels (i.e., *$p<.05$, **$p<.01$, ***$p<.001$). Significant results described in text.}
\begin{tabular}{@{}lllll@{}}
\toprule
\parbox[t]{2in}{\textbf{Impact of personality traits with AI on engagement}} & $adj.R^2=-.015$ &  $p=.726$  &  &  \\
\midrule
Predictor & $B$ & Std. error & $\beta$ & $p$-value \\ 
\midrule
Constant & 4.850 & 1.573 &  &  \\
Openness & -.014 & .231 & -.006 & .951 \\
Conscientiousness & .145 & .195 & .070 & .456 \\
Extraversion & .162 & .176 & .089 & .358 \\
Agreeableness & -.150 & .205 & -.062 & .466 \\
Neuroticism & .047 & .174 & .025 & .788 \\
\textbf{} & \textbf{} & \textbf{} & \textbf{} & \textbf{} \\
\midrule
\parbox[t]{2in}{\textbf{Impact of personality traits with AI on perceived understanding}} & $adj.R^2=.015$ & $p=.210$ &  &  \\
\midrule
Predictor & $B$ & Std. error & $\beta$ & $p$-value \\ 
\midrule
Constant & 2.289 & .452 &  &  \\
Openness & .087 & .067 & .117 & .196 \\
Conscientiousness & .020 & .056 & .033 & .724 \\
Extraversion & .029 & .051 & .054 & .572 \\
Agreeableness & -.003 & .059 & -.004 & .963 \\
Neuroticism & .066 & .050 & .121 & .191 \\
\textbf{} & \textbf{} & \textbf{} & \textbf{} & \textbf{} \\
\midrule
\parbox[t]{2in}{\textbf{Impact of personality traits and age with AI on actual understanding}} & $adj.R^2=.111$ & $\boldsymbol{p=<.001***}$ &  &  \\
\midrule
Predictor & $B$ & Std. error & $\beta$ & $p$-value \\ 
\midrule
Constant & 17.755 & 1.512 &  &  \\
Openness & -.472 & .210 & -.192 & \textbf{.026*} \\
Conscientiousness & -.268 & .181 & -.134 & .142 \\
Extraversion & -.031 & .160 & -.018 & .846 \\
Agreeableness & -.227 & .186 & -.097 & .225 \\
Neuroticism & .076 & .158 & .042 & .630 \\
Age & -.055 & .019 & -.231 & \textbf{.005**} \\
 &  &  &  &  \\
\midrule
\parbox[t]{2in}{\textbf{Impact of personality traits with AI on trust}} & $adj.R^2=.037$ & $p=.063$ &  &  \\
\midrule
Predictor & $B$ & Std. error & $\beta$ & $p$-value \\ 
\midrule
Constant & 20.836 & 2.789 &  &  \\
Openness & -.849 & .410 & -.184 & \textbf{.041*} \\
Conscientiousness & .494 & .345 & .131 & .154 \\
Extraversion & .283 & .312 & .085 & .366 \\
Agreeableness & .681 & .363 & .156 & .063 \\
Neuroticism & -.380 & .309 & -.112 & .220 \\ \bottomrule
\end{tabular}
\label{tab:personality-results}
\end{table*}

Table~\ref{tab:personality-results} shows the results of multiple regressions to test the impact of personality traits and age (previous significant result, Table~\ref{tab:age-gender-prev_exp-results}) on engagement, perceived understanding, actual understanding, and trust. When tested the association between the five personality traits and our participants' engagement with explanations, we found no statistically significant effect. Similarly, a model predicting perceived understanding had no statistically significant effect. 

However, the picture was slightly different for predicting actual understanding of the explanations, with a statistically significant effect. Given that age was found to be significant, we also included it in this model. Out of the five personality traits, only openness had a significantly statistical negative effect, while age was weakly and negatively associated with actual understanding. 

For trust, we again found no statistically significant effect. However, as in the previous model, the trait of openness had a negative effect on the degree of trust our participants attributed to the explanations.

In addition to the previous models, we tested a multiple regression model with all variables combined (i.e., age, gender, previous experience, and the five personality traits) to predict our participants' engagement, perceived understanding, actual understanding, and trust in explanations. 
To assess the absence of multicollinearity among the independent variables, we conducted an analysis using Variance Inflation Factors (VIF) \cite{miles_tolerance_2014}. Typically, VIF values exceeding 5 indicate potential multicollinearity concerns. However, in our study, the highest VIF value calculated for any independent variable was 1.39, comfortably below this threshold, suggesting no significant multicollinearity issues \cite{miles_tolerance_2014}.
Upon analyzing the full model for engagement, we observed an inadequate fit ($F(8,140)=0.419$, $p=.908$, $adj. R^2=-.032$), with none of the independent variables showing significance -- a result consistent with our findings from running the models separately. For perceived understanding, the model marginally fit ($F(8,140)=2.027$, $p=.047$, $adj. R^2=-.053$), with age being the only statistically significant predictor. For actual understanding, the model fit (similar to running the models separately), with age and openness being the only two statistically significant predictors ($F(8,140)=3.107$, $p=.003$, $adj. R^2=-.102$). For trust, the model did not fit compared to running the models separately ($F(8,140)=1.804$, $p=.081$, $adj. R^2=-.042$). Overall, the model for predicting perceived understanding ($adj. R^2 = 0.053$) and trust ($adj. R^2 = 0.042$) had slight differences compared to running the models with a subset of predictors. However, these differences can be considered negligible as both models' predictive power is relatively small. Additionally, out of the four new models, the same set of variables was statistically significant (i.e., age and openness), reinforcing our finding that no other user characteristics had any predictive power.

\section{Discussion}
\label{sec:discussion}
The aim of our study was to investigate how user characteristics, such as gender, age, expertise, and personality traits affect engagement, perceived/actual understanding, and trust of XAI systems. Our findings showed that user characteristics in general influenced these measures very little. We only found one significant effect: a negative effect of both age and openness on actual understanding. While there was a significant effect of openness on trust, the regression model fit was not significant. This suggests that lay users of XAI might be much more similar to each other than they are different. We now discuss the implications of our work in light of the limitations of our study, previous research, and lessons for future research.

\subsection{Limitations}
We acknowledge that there are several aspects that could have been improved in our study. First, the sample size was relatively limited, and despite our power calculations, our statistical tests may not detect very small effects. 

Second, there may have been confounds that introduced noise into our study data, due to using an online study setup. For example, we had no control over the environment in which participants completed the study, their dedication to completing their study, their real engagement with the explanations, etc. 

Third, our study focused specifically on explanations provided for a toxicity classifier. While this is an interesting domain that nonetheless can be easily understood by lay users, previous work \cite{bucinca_proxy_2020} has already pointed out problems with using proxy tasks as compared to real tasks. This might limit the confidence with which results can be taken up. We suggest that the repeatability of our results should be investigated in real decision-making tasks and perhaps high-impact domains. 

Fourth, our research focused primarily on correlations and statistical testing using multiple regression models due to the specific nature of our research questions. However, this methodological approach diverged from the analysis methods employed in previous studies (e.g., \cite{millecamp_whats_2020, conati_toward_2021}) investigating the impact of personality on XAI in various applications such as music recommendation and tutoring systems. As a consequence, a direct comparison of our results with these studies was impossible. Our choice of analysis methodology was influenced partly by the size of the available data and the limited variables that could be controlled in our study. In future investigations encompassing a broader range of variables and larger-scale data, we intend to move beyond correlation and regressions and perform more nuanced analysis (e.g. causal analysis \cite{pearl_causality_2009}).

Finally, our study did not differentiate between types of explanation content or presentation. Hence, we can only confirm the impact of user characteristics on explanations \emph{as a whole}, not reactions to specific components of explanations. Similarly, we only evaluated one type of explanation generation approach, LIME \cite{ribeiro_why_2016}, and thus cannot speak to the effects of utilizing other explanations approaches, e.g. SHAP \cite{lundberg_unified_2017} or Integrated Gradients (IG) \cite{sundararajan_axiomatic_2017}. In addition, our study only examined one type of explanation, i.e. showing word importance, while other explanation techniques were not investigated (e.g. example-based explanations \cite{sokol_glass-box_2018}).

\subsection{Previous Work on User Characteristics}

Previous work investigating the perception and use of AI and XAI has suggested that user characteristics such as gender, age, previous experience as well as personality traits could be used to personalize experience, e.g. \cite{dhelim_survey_2022, tintarev_survey_2007, reeder_evaluating_2023}. Our research used this related work as a launchpad to address the impact of user characteristics on interacting with explanations of AI, using standard statistical methods as tools. Our findings problematize using user characteristics for personalizing explanations. Specifically, our results indicate that lay users of XAI systems may be more alike than different from one another, corroborating other work that cautions against the use of gender as a factor in preferences for explanations~\cite{tran_user_2023}. 

This raises an important question that both HCI and XAI researchers and practitioners must address: \emph{Is personalization an effective strategy for AI explanations?} In light of our findings, we endeavor to seek answers to this question, and argue that, as a community, we need to (re)think whether personalization for AI explanations may lead to a rabbit hole. Drawing from personalization research in other domains, we posit that researchers in XAI might be subject to the same issues due to: \emph{a)} divergence from fields that study users and their characteristics; \emph{b)} the ease of obtaining user characteristics; \emph{c)} the conceptual ease of applying user characteristics; \emph{d)} precedence, \emph{e)} authority approval. First, other fields have extensively studied human-human relationships and the effects of, for example, personality on trust \cite{hancock_how_2023}. \citet{takami_personality-based_2023} demonstrated that personality-based explanations for e-learning recommender systems were more effective than conventional explanations. However, much of this research and changes to theories and implications never make it back to the research in XAI. Second, it is easy to obtain user characteristics such as age, gender, expertise, and even personality; in fact, most of these are supplied by participant recruitment or crowd-sourcing platforms or can be captured easily in user studies. It is often a matter of ``we have the data so might as well use it''. However, these user characteristics might not be exhaustive. Previous literature suggested that people rarely engage analytically with each individual AI recommendation and explanation, and instead develop general heuristics about whether and when to follow the AI suggestions~\cite{bucinca_trust_2021}; observations that require to tap into people's cognitive abilities (e.g., by capturing it through the Need for Cognition, which is a stable personality trait that captures one's motivation to engage in effortful mental activities~\cite{cacioppo_need_1982}). Third, interactions with XAI, as with any technology, are complex and thus its success is not guaranteed. Thus, we might look for an easy answer to account for why a system did better than another, and thus turn to user characteristics as a cause for differences in user behavior or acceptance. In fact, explanations could inadvertently foster a false sense of security by obscuring the inherent uncertainty of the underlying models. Previous literature suggests a need to (re)consider the target audience when designing XAI~\cite{ehsan_who_2021}, recognizing that individuals with different roles and levels of expertise might interpret explanations in varied ways~\cite{jiang_who_2022}. Fourth, work builds each other; one paper investigating a user characteristic can spawn papers on the same or other user characteristics. Looking back there is a long history of work in HCI on user characteristics and modeling users. Finally, there is also the matter of authority approval. For example, Gunning et al. \cite{gunning_xaiexplainable_2019} suggest that explanations should be targeted to the user which might imply that a green light is given to investigate user characteristics in XAI.

\subsection{Future Research Challenges}
Our work points the way to five challenges that need to be addressed in future research in this area.

\subsubsection{Repeatability} So far, there have been many studies that have investigated some aspects of user characteristics and their effects on explanation use, e.g. \cite{reeder_evaluating_2023, tran_user_2023, szymanski_visual_2021}. However, the majority of these pertinent existing studies have not shared the data or code for analysis. While we did not find any strong patterns in our study, there is a need for confirming our results. We, therefore, make our framework -- prototype, code, measures, and data -- available for researchers to replicate our study, and we encourage the XAI research community to do the same to ensure a growing body of evidence that can be systematically reviewed. However, further steps need to be taken to ensure the viability of these investigations, as demonstrated by the next two points.

\subsubsection{Measuring user characteristics and XAI} Recall that we used TIPI to measure personality traits. As pointed out in section 3.4, TIPI is generally comparable with other, longer forms of personality trait instruments but struggles to measure openness robustly. This might explain why we got the results we did: we found that individuals with higher levels of openness had lower levels of trust in XAI systems, a very counter-intuitive result. We believe that this might be due to \emph{measuring errors}, where we possibly measure something else in the users' characteristics entirely. Similarly, we used a common, very granular measure for trust (without accounting for any baseline trusting attitude) and extended previous approaches using `mental model soundness' \cite{kulesza_tell_2012} for measuring actual understanding. Until measuring both user characteristics (as independent variables) and effects (as dependent variables) are standardized within our community, or until others have employed our code and protocol to replicate our measurements, we will not be able to compare results across studies or come to solid conclusions that help us shape theory for the use and design of XAI systems, whether to contemplate personalized XAI based on user characteristics or not.

\subsubsection{Characterization of users}
Our results considered a limited subset of user characteristics, such as age, gender, previous experience with AI, and personality traits. Recent work has also suggested that other user characteristics might matter in how people respond to explanations. For example, \emph{need for cognition} as a personality trait has been investigated as to its role in explanations \cite{bucinca_trust_2021,millecamp_explain_2019,millecamp_whats_2020} and there are some promising results in this direction. Other user characteristics beyond gender, age, or previous experience might also factor into how explanations are perceived or used. For example, \emph{information processing style} which has been linked to gender \cite{burnett_gendermag_2016} might influence how explanations are interpreted. It might also be that \emph{graph interpretation competence} \cite{glazer_challenges_2011} can play a role, especially given the visual nature of many explanation approaches. Future work is warranted to uncover other user characteristics that remain stable and provide robust results.

\subsubsection{Personalization in practice}
Our results suggest that incorporating users' age and, maybe, openness (but see our previous discussion about the limitations of this dimension of personality), into the design and implementation of XAI systems could be beneficial for increasing individuals' understanding of these systems. This raises a practical question of how to do that. Are we expecting users to fill in a possibly lengthy questionnaire every time they use an XAI system? Building a user profile only seems worthwhile for the sustained use of a system that provides explanations, where the benefit of personalized design outweighs the initial cost of providing information. Currently, there are very few XAI systems targeted at lay users that are to be used repeatedly.
In addition, we still do not know how to design explanations to take user characteristics into account. For example, what designs targeted at older users might increase their understanding? Further research into specific components of explanations might yield some answers in the future. 

\subsubsection{The value of good interaction design}

Finally, maybe we are going about this in the wrong way entirely, at this point in the research landscape. In our study, we focused on a relatively static presentation of explanations (highlighting word importance) for all users, minimizing the necessity for extensive user interaction. We do know that different stakeholders might have diverging needs of explanations that a system provides and that the content might need to differ in response. Similarly, we might need to adjust the presentation of explanations in light of stakeholders' needs. It has been suggested \cite{gunning_xaiexplainable_2019} that these stakeholders could be segmented into user groups (e.g., data scientists, domain experts, regulators/auditors), and that these should be considered in the explanation design. We argue that at this stage we are jumping ahead too far, without having achieved a solid grounding for addressing the requirements of these larger user groups. We encourage other researchers to investigate differences between stakeholder groups in order to build better, more targeted XAI systems.

\section{Reproducibility}
To allow for reproducibility, we made our data and code publicly available: \url{https://github.com/RobertNimmo26/Toxic-Comments-XAI-Study}

\section{Author Positionality Statement}
It is important to acknowledge that our backgrounds and experiences may have shaped our positionality~\cite{frluckaj_gender_2022, havens_situated_2020}. Within the context of this work, we situate ourselves in the United Kingdom during the 21\textsuperscript{st} century, writing as authors primarily engaged in academic and industry research. Our research team comprises one female and four male individuals originating from the United Kingdom, Southern Europe, and East Asia, with diverse ethnic and religious backgrounds. Our collective expertise spans various fields, including artificial intelligence, explainable AI, human-computer interaction, computational social science, ubiquitous computing, and software engineering.

\section{Conclusion}
\label{sec:conclusion}

In this paper, we explored the impact of user characteristics on XAI. We ran an online study where participants checked a toxicity classifier that provided explanations and then examined the relationship between user characteristics of gender, age, previous experience, and personality on engagement, perceived and actual understanding, and trust. We found that the users' characteristics we collected during the study did not have a significant effect on most outcome measures. Furthermore, the only significant effect was a negative relationship between both age and openness to participants' actual understanding of the XAI system. Overall, our work sheds light on employing user characteristics of XAI research, and our discussion problematizes the direction of future research in this area. Our study is evidence that lay users may be more similar to each other than different and that XAI designs should possibly concentrate on more marked differences between larger user groups. 

\bibliographystyle{ACM-Reference-Format}
\bibliography{main}

\end{document}